\documentclass[fleqn,11pt]{article}

\usepackage{caption}
\usepackage{mathptmx}
\usepackage{url}
\usepackage{parskip}
\usepackage{paralist}
\usepackage{multirow}
\usepackage{booktabs}
\usepackage{amsmath, amscd, amsfonts, amssymb, graphicx, color}
\usepackage{hyperref}
\usepackage{cite}
\usepackage{epsfig}
\usepackage{amsthm}
\usepackage{epstopdf}
\usepackage{footnote}
\usepackage{newlfont}

\begin{document}
%%%%%%%%%%%%%%%%%%%%%%%%%%%%%%%%%%%%%%%%%%%%%%%%%%%%%%%%

 \title{A new approach for image compression using normal matrices}

 \author{E. Kokabifar\thanks{Department of Mathematics,
 Faculty of Science, Yazd University, Yazd, Iran
 (e.kokabifar@stu.yazd.ac.ir, loghmani@yazd.ac.ir).},\,
 G.B. Loghmani\footnotemark[1]\,\,
 and  A. Latif\thanks{Department of Electrical and Computer Engineering,
  Yazd University, Yazd, Iran (alatif@yazd.ac.ir).}}
\maketitle

\vspace{-6mm}

\begin{abstract}
In this paper, we present methods for image compression on the basis of eigenvalue decomposition of normal matrices. The proposed methods are convenient and self-explanatory, requiring fewer and easier computations as compared to some existing methods. Through the proposed techniques, the image is transformed to the space of normal matrices. Then, the properties of spectral decomposition are dealt with to obtain compressed images. Experimental results are provided to illustrate the validity of the methods.
\end{abstract}


%%%%%%%%%%%%%%%%%%%%%%%%%%%%%%%%%%%%%%%%%%%%%%%%%%%%%%%%%%%%%%%%%%%%%%%
\section{Introduction}\label{intro}
Nowadays, digital images and other multimedia files can become very large in size and, therefore, occupy a lot of storage space. In addition, owing to their size, it takes more time to move them from place to place and a larger bandwidth to download and upload them on the Internet. So, digital images may pose problems if we regard the storage space as well as file sharing. To tackle this problem, \textit{image compression} which deals with reducing the size of an image (or any other multimedia) file can be used. Image compression actually refers to the reduction of the amount of image data (bits) required for representing a digital image without causing any major degradation of the image quality. By eliminating redundant data and efficiently optimizing the contents of a file image, provided that as much basic meaning as possible is preserved, image compression techniques, make image files smaller and more feasible to share and store.

%Now let us for the subject to be fully grasped, present certain basic definitions seems necessary. A \textit{gray scale image} can be demonstrated as a matrix which the magnitude of each of whose components is called \textit{gray level} or \textit{intensity} of the image at that point. A given image is called a digital image if all of its elements are finite and discrete quantities; in this case, these elements are called \textit{pixels}. Digital image compression can be construed as a \textit{Digital Image Processing} (DIP) which concerns using digital computers to preform processing on digital images. DIP receives a digital image as an input, processes efficient algorithms on it by a computer and gives an image as an output. 
The study of digital image compression has a long history and has received a great deal of attention especially with respect to its many important applications. References for theory and practice of this method are \cite{ref1,ref2,ref3,ref4,ref5} and \cite{ref6}, to name but a few.

Image compression, as well as other various fields of digital image processing, benefits from the theory of linear algebra as a helpful tool. In particular, Singular Value Decomposition (SVD) is one of the most useful tools for image compression \cite{svd1,svd2}.

The matrix $A \in \mathbb{C}^{m\times n}$ can be written in the form of $USV^{*}$, where $U\in\mathbb{C}^{m\times m}$ and $V\in\mathbb{C}^{n\times n}$ are unitary matrices, i.e., $U^*U=I_m, V^*V=I_n$, where $^*$ denotes complex conjugate transpose and $I_n$ is $n \times n$ identity matrix. The matrix $S$ is an $m\times n$ diagonal matrix in such a way that its nonnegative entries are ordered in a non-increasing order (see for example, Theorem 7.3.5 of \cite{horn}). With respect to the influences of singular values of $A$ in compressing an image, and considering the important point that the singular values of $A$ are the positive square roots of the eigenvalues of matrices $A^*A$ and $AA^*$, the present study concerns itself with the eigenvalue of the normal matrices $A+A^*$ and $A-A^*$ on the purpose of establishing certain techniques for image compression that are efficient, lead to desirable results and need fewer calculations.

In the next section, we briefly present some definitions and concepts about normal matrices. Section \ref{comp} consists of two subsections in which the proposed image compression methods are explained. In Section \ref{exp}, the validity rates of the presented image compression schemes are investigated and compare their efficiencies by experimental results.

\section{Normal matrices} \label{normal}
In this section, we review the definition and some properties of normal matrices. See \cite{normal1,normal2} and the references mentioned there as the suggested sources on a series of conditions on normal matrices. In the next section, we will describe the proposed method on the basis of these presented properties.

A matrix $M\in\mathbb{C}^{n \times n}$ is called \textit{normal} if $M^*M=MM^*$. Assuming $M$ as an $n$-square normal matrix, there exists an orthonormal basis of $\mathbb{C}^{n\times n}$ that consists of eigenvectors of $M$, and $M$ is unitarily diagonalizable. That is, let the scalars $\lambda_1,\ldots, \lambda_n$, counted according to multiplicity, be  eigenvalues of the normal matrix $M$ and let $u_1, \ldots, u_n$ be its corresponding orthonormal eigenvectors. Then, the matrix $M$ can be factored as the following:
\begin{equation*}
M=U\Lambda U^*=
\sum\limits_{i = 1}^n {\lambda _i u_i u_i^* } ,\qquad \Lambda=\mbox{diag}(\lambda_1,\ldots,\lambda_n),\qquad U=[u_1,\ldots,u_n],
\end{equation*}
where the matrix $U$ satisfies $UU^*=I_n$. Maintaining the generality, assume that eigenvalues are ordered in a non-ascending sequence of magnitude, i.e., $\left| {\lambda _1 } \right| \ge \left| {\lambda _2 } \right| \ldots  \ge \left| {\lambda _n } \right|$.

It is to be noticed that, if all the elements of the matrix $M$ are real, then $M^*=M^T$, where $M^T$ refers to the transpose of the matrix $M$.  A square matrix $M$ is called \textit{symmetric} if $M=M^T$ and called \textit{skew-symmetric} if $M=-M^T$. That symmetric and skew-symmetric matrices are normal is easy to see. Also, the whole set of the eigenvalues of a real symmetric matrix are real, but all the eigenvalues of a real skew-symmetric matrix are purely imaginary.
A general square matrix $M$ satisfies $M=B+C$, for which the symmetric matrix $B=(M+M^T)/2$ is called the \textit{symmetric part} of $M$ and, similarly, the skew-symmetric matrix $C=(M-M^T)/2$ is called the \textit{skew-symmetric} part of $M$. As a consequence, every square matrix may be written as the sum of two normal matrices: a symmetric matrix and a skew-symmetric one. We specially use this point in the proposed image compression techniques.

\section{Image compression methods}\label{comp}
This section consists of two subsections where methods for image compression are presented using normal matrices. To this purpose, the matrix representing the image is transformed into the space of normal matrices. Next, the properties of its eigenvalue decomposition are utilized, and some less significant image data are deleted. Finally, by returning to the original space, the compressed image can be constructed.

Let $X$ be an $n\times n$ matrix to represent the image. Two distinct methods are taken into account. First, the symmetric parts of $X$ are dealt with to establish an image compression scheme. This procedure can be performed in the same way for the skew-symmetric parts of the matrix $X$. Next, another technique is explained using both symmetric and skew-symmetric parts of the matrix $X$. What is noticeable is that finding the eigenvalues and eigenvectors of a matrix requires fewer calculations than finding its singular values and singular vectors. Moreover, it is possible to calculate the eigenvalues and eigenvectors of a normal (especially symmetric or skew-symmetric) matrix by explicit formulas and, therefore, may yet again need less computation \cite{sym1,sym2,sym4}.

\subsection{Image compression method using the symmetric part of $X$}\label{sb1}
In this subsection, a technique of image compression comes into focus on the basis of the eigenvalue decomposition of the symmetric parts of the matrix $X$. This can be performed in the same way for the skew-symmetric parts of $X$. It is to be noted that, for the same image, the results obtained by these two techniques (i.e. using symmetric or skew-symmetric part) may be different.

Assume $B_X$ as the symmetric part of the matrix $X$. The normal matrix $B_X$ can be factored as in the following:
\[
B_X  = U_{B_X } \Lambda _{B_X } U_{B_X }^*=
\sum\limits_{i = 1}^n {\lambda _{B_{X ,i}} u_{B_{X ,i}} u_{B_{X ,i}}^* }  ,\qquad\Lambda _{B_X }  = \mbox{diag}(\lambda _{B_{X ,1}} , \ldots ,\lambda _{B_{X ,n}} ),
\]

Now, bearing in mind that the eigenvalues are sequenced in a non-ascending order of magnitude, compress the symmetric part of the image by wiping off the small enough eigenvalues of $B_X$. If $k$ of the larger eigenvalues remains, then there is
\begin{equation}\label{sym}
\tilde{B}_X=\sum\limits_{i = 1}^k {\lambda _{B_{X ,i}} u_{B_{X ,i}} u_{B_{X ,i}}^* } \;; \qquad k\le n,
\end{equation}
where the total storage for $\tilde{B}_X$ is $k(n+1)$.
Here comes the stage of reconstructing the compressed image $\tilde{X}$ from its symmetric part. To this purpose, all the elements located above the main diagonal of the matrix $X$ are needed, and this calls for ${{n\left( {n - 1} \right)}}/2$ storage spaces. Because $\tilde{X}$ can be considered as an acceptable approximation of $X$, let us take all the elements above the main diagonal of $\tilde{X}$ as the elements of $X$. Obviously, the elements located below the main diagonal of the matrix $\tilde{X}^T$ should be determined too. In addition, it is inferred from the fact $2\tilde{B}_X=\tilde{X}+\tilde{X}^T$ that the elements located below the main diagonal of $\tilde{X}$ can be obtained by subtracting the elements below the main diagonal of $2\tilde{B}_X$ from the elements of $\tilde{X}^T$ . See the following equation where $"\checkmark"$ and $"\times"$ denote the given and unknown entries, respectively.
\[
2\tilde{B}_X  = \mathop {\left[ {\begin{array}{*{20}c}
  \times & {} & \checkmark  \\
   {} &  \ddots  & {}  \\
   \times & {} & \times  \\
\end{array}} \right]}\limits_{\tilde{X}}  + \mathop {\left[ {\begin{array}{*{20}c}
   \times & {} & \times  \\
   {} &  \ddots  & {}  \\
   \checkmark & {} & \times  \\
\end{array}} \right]}\limits_{\tilde{X}^T }.
\]

It is clear that the main diagonal elements of $\tilde{X}$ are the same as those on the main diagonal of $\tilde{B}_X$.
It is also to be noted that, by this procedure, only the elements located below the main diagonal of $X$ are modified, and its other elements are remain untouched.

Moreover, to reconstruct the compressed image $\tilde{X}$, the elements located below the main diagonal of $X$ may be reserved instead of those above the main diagonal, and then a procedure similar to what has been performed newly is to be followed. Indeed, $X$ can be partitioned into several segments, and those segments may be reserved provided that $\tilde{X}$ is reconstructed. This may be useful specially when some segments of the image have more significance or unchanging (or uncompressing) some partitions of the image is desirable during the image compression.

Finally, it should be pointed out that reconstruction of the compressed image $\tilde{X}$ requires $\left(k(n+1)+\frac{{n\left( {n - 1} \right)}}{2}\right)$ storage spaces when the symmetric part of $X$ is dealt with. However, if the skew-symmetric part of $X$ is concerned, the diagonal elements of the $\tilde{X}$ cannot be obtained from $\tilde{B}_X$ and, therefore, they must be reserved. Reconstructing the compressed image $\tilde{X}$, thus, requires $\left(k(n+1)+\frac{{n\left( {n + 1} \right)}}{2}\right)$ storage spaces when the skew-symmetric part of $X$ is considered for the image compression method. In the next subsection, another image compression technique is provided, for which reconstructing the compressed image demands fewer storage spaces. 
% % % % % % % % % % % % % % % % % % % % % % % % % % % % % % % % % % % % % % % % % % % % % %
\subsection{Image compression method using both symmetric and skew-symmetric parts of $X$}\label{sb2}

The previous subsection introduced an image compression scheme dealing with almost half of an image. By that technique, more than half of the elements of the image remained unchanged. This may be of some usages and advantages. Compressed images obtained by the presented method have a high quality. However, compressing of just half of the original image may cause the method to lose its reliability as compared to some other image compression schemes. To tackle this problem, a new method is presented here about both symmetric and skew-symmetric parts of the matrix $X$ in order to compress the image all over. This new image compression technique is found to be of a remarkably high reliability.

As previously mentioned, every matrix equals the sum of its symmetric and skew-symmetric parts. To establish the new method, with the definition of $B_X$, borne in mind, $C_X$ is used to represent the skew-symmetric part of $X$. The matrix $C_X$ can be written as follows:
\[
C_X  = U_{C_X } \Lambda _{C_X } U_{C_X }^*=
\sum\limits_{i = 1}^n {\lambda _{C_{X ,i}} u_{C_{X ,i}} u_{C_{X ,i}}^* }  ,\qquad\Lambda _{C_X }  = \mbox{diag}(\lambda _{C_{X ,1}} , \ldots ,\lambda _{C_{X ,n}} ),
\]
With respect to the method described in the previous subsection, for an integer $k\le n$, set
\begin{equation}\label{ssym}
\tilde{C}_X=\sum\limits_{i = 1}^k {\lambda _{B_{X ,i}} u_{C_{X ,i}} u_{C_{X ,i}}^* },
\end{equation}
Through (\ref{sym}) and (\ref{ssym}), the compressed image $\mathcal{X}$ will be $\mathcal{X}=\tilde{B}_X+\tilde{C}_X.$ As in the case of reserving the matrix $\tilde{B}_X$, $k(n+1)$ storage spaces are required for saving the matrix $\tilde{C}_X$. As a result, the total storage requirement for $\mathcal{X}$ is $2k(n+1)$.

% % % % % % % % % % % % % % % % % % % % % % % % % % % % % % % %
\section{Experimental results}\label{exp}
In this section, the validity and the influence of the proposed image compression method are examined. Let us note that the ideas presented in this paper, can readily be used  to establish a block-based compression scheme. This scheme concerns dividing of an image into non-overlapping blocks and compressing of each block.
%The block-based compression method using SVD is especially well known (e.g., see \cite{svdblock}). 

%In our experiments, the compression method explained in Subsection \ref{sb1} is considered for using both the symmetric and skew-symmetric parts of the matrix representing the image. Also, the compression scheme demonstrated in Subsection \ref{sb2} has been tested to prove its efficiency.

The Peak Signal to Noise Ratio (PSNR) is calculated to measure the quality of the compressed image. In the case of gray scale images of size $M\times N$, whose pixels are represented with 8 bits, PSNR is computed as follows:
\[
PSNR = 10\log _{10}^{\frac{{255^2 }}{{MSE}}} ;\qquad MSE = \frac{1}{{MN}}\sum\limits_{i,j}^{} {\left| {X_{i,j}  - \mathcal{X}_{i,j} } \right|^2 },
\]
where $X_{i,j}$ and $\mathcal{X}_{i,j}$ refer to the elements of the original and the compressed images respectively. In the above relationship, MSE stands for the Mean Square Error between the original image and the compressed image pixels. In addition, Compression Ratio (CR) may be calculated as an important index to evaluate how much of an image is compressed. CR is the amount of bits in the original image divided by the amount of bits in the compressed image; that is, 
\[
\mbox{CR} = \frac{\mbox{Original Image Size}}{\mbox{Compressed Image Size}}.
\]
For the sake of simplicity, Method \#1 and Method \#2 are branded as image compression techniques which use the symmetric and the skew-symmetric parts of the matrix representing the image, in the image compression schemes, respectively. In addition, let Method \#3 denominate the image compression technique described in Subsection \ref{sb2} and let Method \#4 stand for the image compression method using SVD. Consequently, the following relationships emerge for CR of these methods: 

\begin{eqnarray*}
&& CR_{Method \#1}  = \frac{{n^2 }}{{\left( {k(n + 1) + \frac{{n(n - 1)}}{2}} \right)}},\qquad CR_{Method \#2}  = \frac{{n^2 }}{{\left( {k(n + 1) + \frac{{n(n + 1)}}{2}} \right)}},\\ && CR_{Method \#3}  = \frac{{n^2 }}{{2k(n + 1)}},\hspace{.9cm} \mbox{and}\hspace{.9cm} CR_{Method \#4}  = \frac{{n^2 }}{{k(2n + 1)}}.
\end{eqnarray*}

\begin{figure}
\centering
\includegraphics[width=0.7\linewidth]{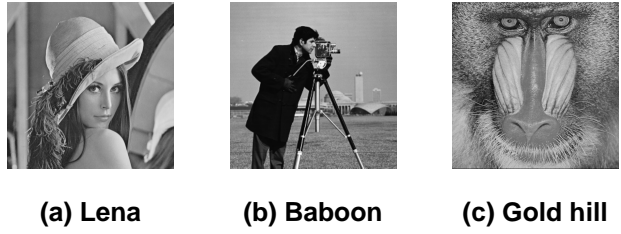}
\caption{Original Images}
\label{fig:images1}
\end{figure}

\begin{figure}
\centering
\includegraphics[width=0.7\linewidth]{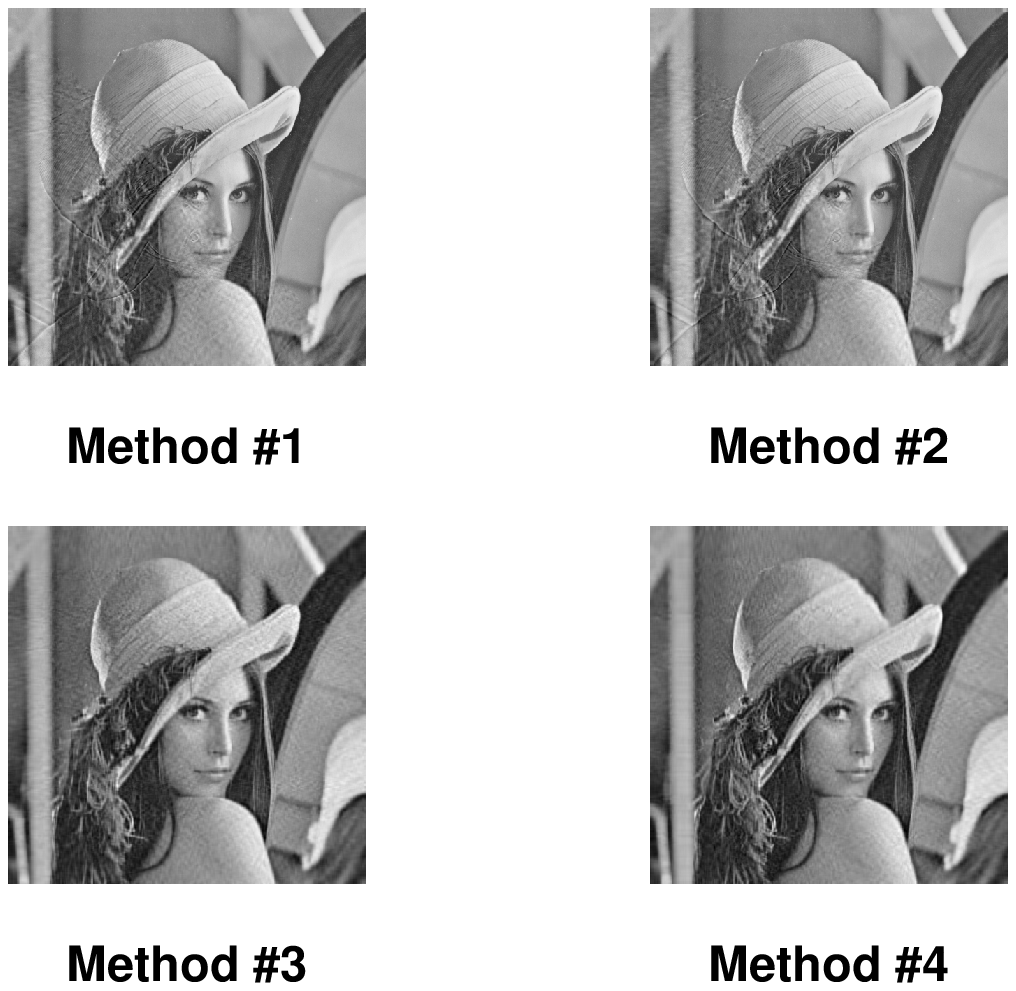}
\caption{Compressed image Lena for $k=50$.}
\label{fig:compressed50}
\end{figure}

\begin{figure}
\centering
\includegraphics[width=0.7\linewidth]{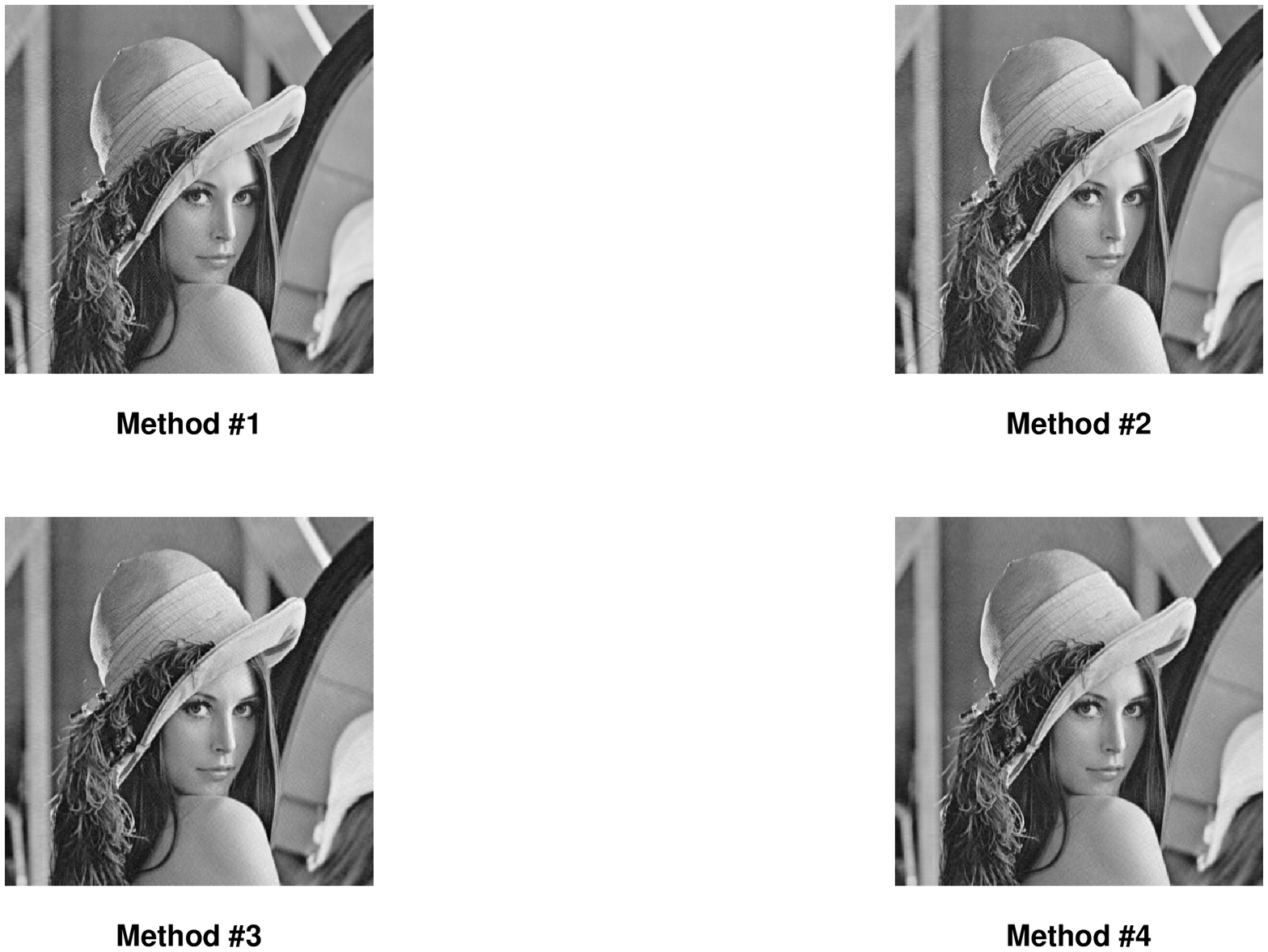}
\caption{Compressed image Lena for $k=100$.}
\label{fig:compressed100}
\end{figure}

%It should be noted that, we can remove all eigenvalues which their magnitudes are smaller than or equal to an arbitrary tolerance and obtain the values of $k$ as the number of remained eigenvalues. We applied this point in our experiments for each image compression techniques.
In the experiments conducted in this study, three $512 \times 512$ gray scale images were considered, including (a) Lena, (b) Baboon and (c) Gold hill presented in Fig \ref{fig:images1}.
The PSNR results are shown in Tables \ref{table:lenap}, \ref{table:baboonp} and \ref{table:goldhillp}, for some integer values of $k$, for images Lena, Baboon and Gold hill, respectively. Also, the CR results are given in Table \ref{table:goldhillc} for a $512 \times 512$ image. The results obtained by these techniques are compared to those achieved by Method \#4 in our tables. Furthermore, Figures \ref{fig:compressed50} and \ref{fig:compressed100} show the compressed image Lena obtained by the proposed techniques as well as image compression method using SVD, Method \#4, for $k=50$ and $k=100$, respectively.
% % % % % % % % % % % % % % % % % % % % % % % % % % % % % % % % % % % %
\begin{table}[ht]
\caption{PSNR results for Lena} % title of Table
\centering % used for centering table
\begin{tabular}{c c c c c} % centered columns (5 columns)
\\ [-2 ex]
\hline\hline %inserts double horizontal lines
 $k$ & Method\#1 & Method\#2 & Method\#3 & Method \#4 \\ [.5ex] % inserts table
%heading
\hline % inserts single horizontal line
\\ [-2 ex]
   10 &   21.7012 & 22.3391 & 22.0050 & 22.4065\\
   20 &   24.8722 & 25.0305 & 24.9462 & 25.2937\\
   30 &   26.8832 & 27.0315 & 26.9531 & 27.2243\\
   40 &   28.3856 & 28.5746 & 28.4751 & 28.7927\\
   50 &   29.7358 & 29.9001 & 29.8129 & 30.1761\\
   75 &   32.5608 & 32.8031 & 32.6763 & 33.1093\\
  100 &   35.0504 & 35.1668 & 35.1047 & 35.6641\\
  150 &   39.2993 & 39.2955 & 39.2938 & 39.8988\\
[.5ex] % [1ex] adds vertical space
\hline %inserts single line
\end{tabular}
\label{table:lenap} % is used to refer this table in the text
\end{table}
% % % % % % % % % % % % % % % % % % % % % % % % % % % % % % % % % % % %
\begin{table}[ht]
\caption{PSNR results for Baboon} % title of Table
\centering % used for centering table
\begin{tabular}{c c c c c} % centered columns (5 columns)
\\ [-2 ex]
\hline\hline %inserts double horizontal lines
 $k$ & Method\#1 & Method\#2 & Method\#3 & Method \#4 \\ [.5ex] % inserts table
%heading
\hline % inserts single horizontal line
\\ [-2 ex]
   10 & 19.7071 & 19.7560 & 19.7270 & 19.8493\\
   20 & 20.6441 & 20.7135 & 20.6736 & 20.7510\\
   30 & 21.3880 & 21.4336 & 21.4053 & 21.5203\\
   40 & 22.0773 & 22.1009 & 22.0842 & 22.2453\\
   50 & 22.7319 & 22.7355 & 22.7286 & 22.9306\\
   75 & 24.2968 & 24.3207 & 24.3042 & 24.5787\\
  100 & 25.8111 & 25.7420 & 25.7723 & 26.1761\\
  150 & 28.7678 & 28.6798 & 28.7194 & 29.3173\\
[.5ex] % [1ex] adds vertical space
\hline %inserts single line
\end{tabular}
\label{table:baboonp} % is used to refer this table in the text
\end{table}
% % % % % % % % % % % % % % % % % % % % % % % % % % % % % % % % % % % %
\begin{table}[ht]
\caption{PSNR results for Gold hill} % title of Table
\centering % used for centering table
\begin{tabular}{c c c c c} % centered columns (5 columns)
\\ [-2 ex]
\hline\hline %inserts double horizontal lines
 $k$ & Method\#1 & Method\#2 & Method\#3 & Method \#4 \\ [.5ex] % inserts table
%heading
\hline % inserts single horizontal line
\\ [-2 ex]
   10&   23.9494 & 23.9081 & 23.9223 & 24.1270\\
   20&   26.3107 & 26.2783 & 26.2884 & 26.6046\\
   30&   27.8179 & 27.8131 & 27.8092 & 28.1740\\
   40&   29.0044 & 28.9622 & 28.9768 & 29.3828\\
   50&   30.0126 & 29.9975 & 29.9978 & 30.4103\\
   75&   32.1111 & 32.1537 & 32.1254 & 32.6597\\
  100&   33.9084 & 33.8903 & 33.8927 & 34.5614\\
  150&   37.0963 & 37.0375 & 37.0626 & 37.8911\\
[.5ex] % [1ex] adds vertical space
\hline %inserts single line
\end{tabular}
\label{table:goldhillp} % is used to refer this table in the text
\end{table}
% % % % % % % % % % % % % % % % % % % % % % % % % % % % % % % % % % % %
\begin{table}[ht]
\caption{CR results for an image of size $512\times 512$} % title of Table
\centering % used for centering table
\begin{tabular}{c c c c c} % centered columns (5 columns)
\\ [-2 ex]
\hline\hline %inserts double horizontal lines
 $k$ & Method\#1 & Method\#2 & Method\#3 & Method \#4 \\ [.5ex] % inserts table
%heading
\hline % inserts single horizontal line
\\ [-2 ex]
10  &    1.9283 &  1.9211 & 25.5501 & 25.5750\\
20  &    1.8582 &  1.8515 & 12.7750 & 12.7875\\
30  &    1.7930 &  1.7867 &  8.5167 &  8.5250\\
40  &    1.7322 &  1.7264 &  6.3875 &  6.3938\\
50  &    1.6754 &  1.6699 &  5.1100 &  5.1150\\
75  &    1.5485 &  1.5438 &  3.4067 &  3.4100\\
100 &    1.4394 &  1.4354 &  2.5550 &  2.5575\\
150 &    1.2617 &  1.2586 &  1.7033 &  1.7050\\   
[.5ex] % [1ex] adds vertical space
\hline %inserts single line
\end{tabular}
\label{table:goldhillc} % is used to refer this table in the text
\end{table}
% % % % % % % % % % % % % % % % % % % % % % % % % % % % % % % % % % % % % % % % %
% % % % % % % % % % % % % % % % % % % % % % % % % % % % % % % % % % % % % % % % %

\section{Conclusion}\label{conc}
In this paper, the eigenvalue decomposition of symmetric and skew-symmetric matrices, as two important kinds of normal matrices, and their properties were applied to obtain image compression schemes. The proposed method is straightforward and uncomplicated ones requiring clear and fewer computations as compared to some exiting methods. This image compression method is the first one ever introduced and concerns the symmetric part of an image, which can also be applicable in the case of the skew-symmetric part of the image. The method is capable of keeping half (or some selected segments) of an image unchanged, and this feature may be of some usages. However, as observed in this study, just half of the original image could be changed, which causes the compressed image to lose its reliability. This is what makes the technique different from some other image compression techniques. Hence, the second image compression scheme using both symmetric and skew-symmetric parts of the original image was proposed. The experimental results show the high reliability of this method. Finally, it is to be pointed out that the proposed method can be used to devise block-based compression techniques.

%%%%%%%%%%%%%%%%%%%%%%%%%%%%%%%%%%%%%%%%%%%%%%%%%%%%%%%%
\end{document}